\documentclass[12pt]{iopart}

\usepackage{iopams}  
\usepackage{setstack}
\usepackage{enumerate}

\usepackage{graphicx}
\usepackage{hhline}
\usepackage{dcolumn}
\usepackage{bm}
\usepackage[dvips]{color}

\begin{document}

\title{Robustness of networks against propagating attacks under vaccination strategies}

\author{Takehisa Hasegawa$^1$ and Naoki Masuda$^{1,2}$}
\address{$^1$ Department of Mathematical Informatics,
The University of Tokyo,
7-3-1, Hongo, Bunkyo-ku, Tokyo, 113-8656, JAPAN.
}
\address{$^2$ 
PRESTO, Japan Science and Technology Agency,
4-1-8 Honcho, Kawaguchi, Saitama 332-0012, JAPAN
}
\ead{hasegawa@stat.t.u-tokyo.ac.jp} 
\ead{masuda@mist.i.u-tokyo.ac.jp}

\begin{abstract}
We study the effect of vaccination on 
robustness of networks against propagating attacks
that obey the susceptible-infected-removed model.
By extending the generating function formalism developed by
Newman (2005), we analytically determine the robustness of
networks that depends on the vaccination parameters.
We consider the random defense where nodes are vaccinated randomly
and the degree-based defense where hubs are preferentially vaccinated.
We show that when vaccines are inefficient,
the random graph is more robust against propagating attacks than the scale-free network.
When vaccines are relatively efficient,
the scale-free network with the degree-based defense is more robust 
than the random graph with the random defense 
and the scale-free network with the random defense.
\end{abstract}

\maketitle

\section{Introduction}

Many real networks, such as the WWW, the Internet, social and biological networks,
have complex connectivity.
A property shared by many types of networks
is the scale-free (SF) degree distribution 
$p_k \propto k^{-\gamma}$,
where $k$ is degree (number of links connected to a node) 
and $p_k$ is the fraction of nodes with degree $k$~\cite{barabasi1999}.
Functions of networks crucially depend on the degree distribution and other structural properties of networks~\cite{albert2002statistical,newman2003structure,boccaletti2006report,dorogovtsev2008critical,barrat2008dynamical,newman2010networks,cohen2010complex}. 

An important property of networks is 
robustness against the removal of nodes caused by failures or intentional attacks.
Albert {\it et al}. studied the robustness of networks against two types of attacks:
random failure, where nodes are sequentially removed with equal probability,
and intentional attack, 
where hubs (i.e., nodes with large degrees) are preferentially removed~\cite{albert_error_2000}.
They showed that SF networks having $\gamma \le 3$
are highly robust against the random failure in the sense
that almost all nodes have to be removed 
to disintegrate a SF network.
However, SF networks are
fragile to the intentional attack in the sense
that the network is destroyed if a small fraction of hubs are removed.
These results have been analytically established~\cite{callaway2000network,cohen2000resilience,cohen2001PRL}.
The robustness of networks against other percolation-like processes such as
the betweenness-based attack~\cite{holme2002attack} and
degree-weighted attacks~\cite{gallos_stability_2005}
have also been studied.

In fact, attacks to networks may occur as propagating processes.
Examples include computer viruses~\cite{Kephart91,Kephart93,Pastor-Satorras2001epidemicPRL}.
The (first) critical infection rate above which a global outbreak occurs
has been obtained for propagating processes such as
the susceptible-infected-removed (SIR)
and susceptible-infected-susceptible (SIS) models.
In particular, for SF networks with $\gamma \le 3$, 
any positive infection rate induces a global outbreak~\cite{albert2002statistical,newman2003structure,boccaletti2006report,dorogovtsev2008critical,barrat2008dynamical,newman2010networks,cohen2010complex,Pastor-Satorras2001epidemicPRL,Pastor-Satorras2001epidemicPRE,
moreno2002epidemic} 
(also see a recent paper~\cite{Castellano2010PRL} in which it is shown that the critical infection rate of the SIS model on SF networks vanishes regardless of the value of $\gamma$).
Another (second) critical infection rate, 
above which the network remaining after
all the infected nodes are deleted (we call it residual network)
is disintegrated,
also exists and is larger than the first critical infection rate.
The residual network is important because a 
second epidemic spread may occur on it~\cite{Newman-Threshold-2005PRL,bansal_impact_2009,Bansal2010Plos,Funk2010PRE}.
By using the generating functions,
Newman derived the second critical infection rate 
for uncorrelated networks
to show that the second critical infection rate is positive
even for $\gamma \le 3$~\cite{Newman-Threshold-2005PRL}. 
In particular, the global outbreak of a second epidemic
requires a larger infection rate
than that of a first epidemic~\cite{Newman-Threshold-2005PRL}.

Defense (i.e., immunization) 
strategies for networks to contain epidemics are also 
of practical importance. Epidemics can be efficiently suppressed 
by various defense strategies such as
target immunization~\cite{Pastor-Satorras2002PRE,Madar2004EPJB}, 
acquaintance immunization~\cite{Madar2004EPJB,Cohen2003PRL,Holme2004EPL,Gallos2007PRE}, and 
graph partitioning immunization~\cite{Chen2008PRL}.
Such a defense strategy
raises the first critical infection rate.
In contrast,
we study the effectiveness of a defense strategy 
in enhancing the robustness of networks, 
i.e., increasing the second critical infection rate.
Using the generating functions~\cite{Newman-Threshold-2005PRL,Newman-Spread-2002PRE}, 
we formulate the effects of defense strategies 
on the robustness of networks.
We determine the two critical infection rates for the following three
combinations of networks and defense strategies: 
random graph with the random defense, in which nodes are vaccinated in a random order,
uncorrelated SF network with the random defense, and
uncorrelated SF network with the degree-based defense, in which nodes are vaccinated in
the descending order of the degree.
When vaccines are inefficient,
the random graph is more robust than the SF network 
in the sense that the second critical infection rate is larger.
When vaccines are relatively efficient,
the SF network with the degree-based defense is more robust 
than the other two cases.
We also discuss the optimization of the vaccine allocation under
a trade-off between the number and efficiency of vaccines available.

\section{Model}


\begin{figure}
\begin{center}
\includegraphics[width=15cm]{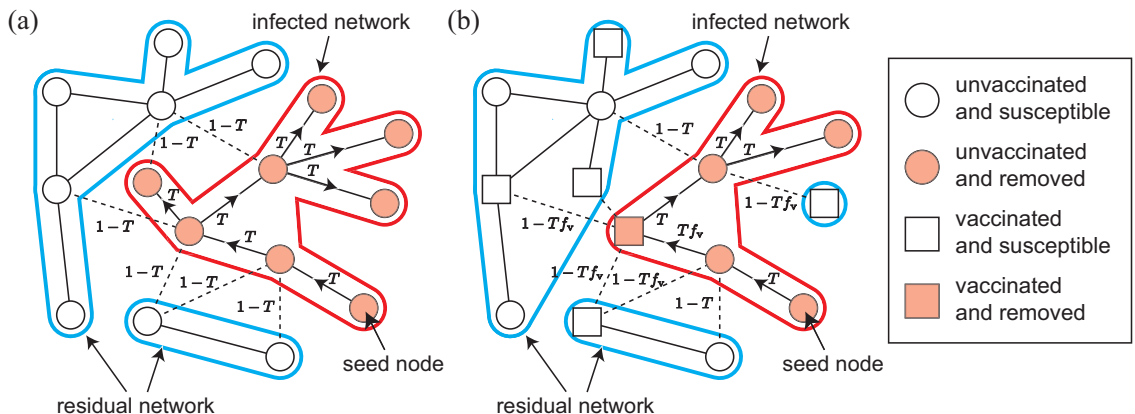}
\end{center}
\caption{
Propagating attack on networks
(a) without and (b) with a defense strategy.
$T$ and $T f_{\rm v}$ are the probabilities that 
an infected (and eventually removed)
node infects an unvaccinated and vaccinated neighbor in a propagating attack, 
respectively.
}
\label{example}
\end{figure}


We adopt the SIR model, which is a continuous-time Markov process on a given static network with $N$ nodes, as the propagating attack. 
Each node takes one of the three states: susceptible, infected, and removed.
All nodes except a randomly selected seed node are initially susceptible.
The seed node is initially infected.
Susceptible nodes get infected at a rate proportional to the number of infected neighbors.
In other words, if a susceptible node is adjacent to an infected node, the susceptible node gets infected with probability $\beta \Delta t$ within short time $\Delta t$. 
An infected node becomes removed at a unit rate,
i.e., with probability $\Delta t$ within short time $\Delta t$, 
irrespective of the neighbors' states.

Following the methods employed in previous literature~\cite{Newman-Spread-2002PRE,grassberger1983on}, 
we analyze the bond percolation to estimate
the transition points, the final size of global outbreaks, and that of residual networks of the SIR model.
We assume that 
the disease is transmitted from an infected node to a susceptible node with probability $T=T(\beta)$, where 
$T=\int [1-\exp(-\beta \tau)] P(\tau)\rmd\tau$, 
$\tau$ is the infectious period during which the infected node is infected, and $P(\tau)$ is the distribution of $\tau$~\cite{Newman-Spread-2002PRE,grassberger1983on}.
In other words, we map the SIR dynamics onto the bond percolation with open bond probability $T$.
Strictly speaking, this mapping is not generally exact~\cite{Kenah2007PRE, Miller2007PRE, KenahandMiller2011}. 
The transition point above which a global outbreak occurs and 
the mean final size of global outbreaks at a given $T$ obtained 
from the analysis of the bond percolation are precise for the SIR model.
However, the bond percolation fails to predict the probability of global outbreaks 
in the SIR model~\cite{Kenah2007PRE, Miller2007PRE, KenahandMiller2011}.
The correspondence between the bond percolation and the SIR model is exact 
if the infectious period in the SIR model is assumed to be constant. 
When analyzing the effects of defense strategies (Sec.~\ref{SecDef}),
we denote 
by $T_i$ the probability that susceptible node $i$ is infected by an infected neighbor.
We set $T_i=T$ for all $i$ unless otherwise stated.

After a propagating attack, 
each node is either susceptible 
or removed (figure~\ref{example}(a)).
We call the network composed
of removed nodes and that composed of susceptible nodes 
infected network and residual network, respectively.
The infected network consists of a single component because 
an attack starts from a single seed.
When $T$ is below a first transition point $T_{\rm c1}$, 
the size of the infected network is $o(N)$, 
whereas the residual network includes a giant component of order $O(N)$.
When $T_{\rm c1} < T < T_{\rm c2}$, where $T_{\rm c2}$ is a second transition point,
both the infected and residual networks have a giant component of order $O(N)$.
When $T>T_{\rm c2}$, 
the giant component of the residual network is absent.

We consider two defense strategies.
In the random defense, $Nf$ randomly selected nodes are vaccinated.
In the degree-based defense, $Nf$ nodes with the largest degrees are vaccinated.
We assume that $T_i$ decreases to
$T_i=T f_{\rm v}$ $(0 \le f_{\rm v} <1)$ for a vaccinated node $i$ 
(figure~\ref{example}(b)).
In epidemiology, the vaccine with $f_{\rm v}=0$ and 
that with $ f_{\rm v} \neq 0$ are called perfect and leaky vaccination, respectively~\cite{bansal_impact_2009,Halloran1992}.

\section{Analysis \label{GFA}}


\begin{figure}
\begin{center}
\includegraphics[width=15cm]{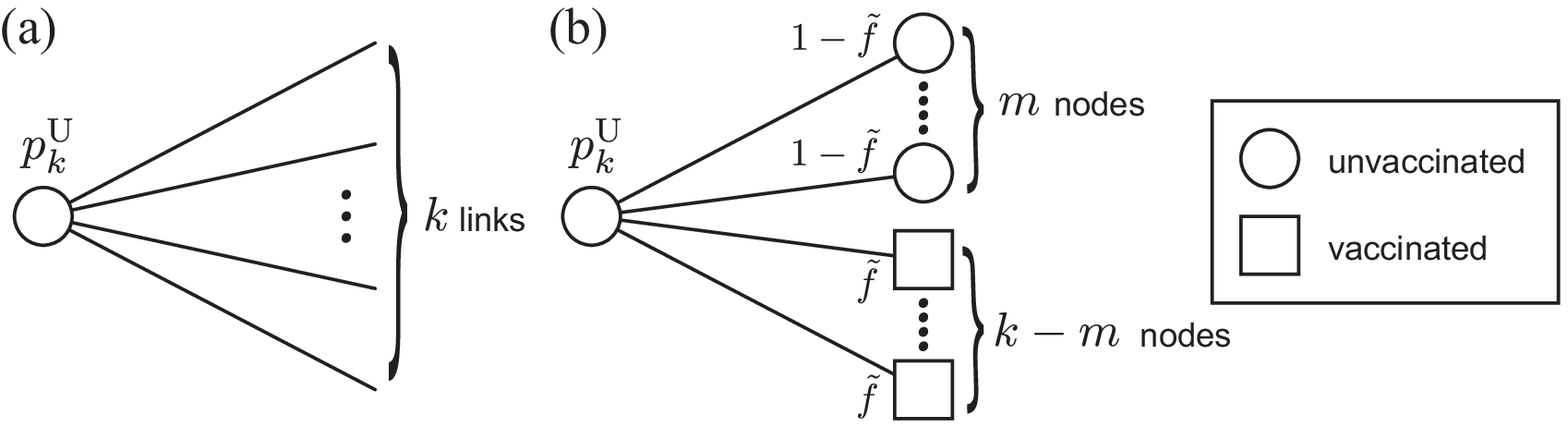}
\includegraphics[width=15cm]{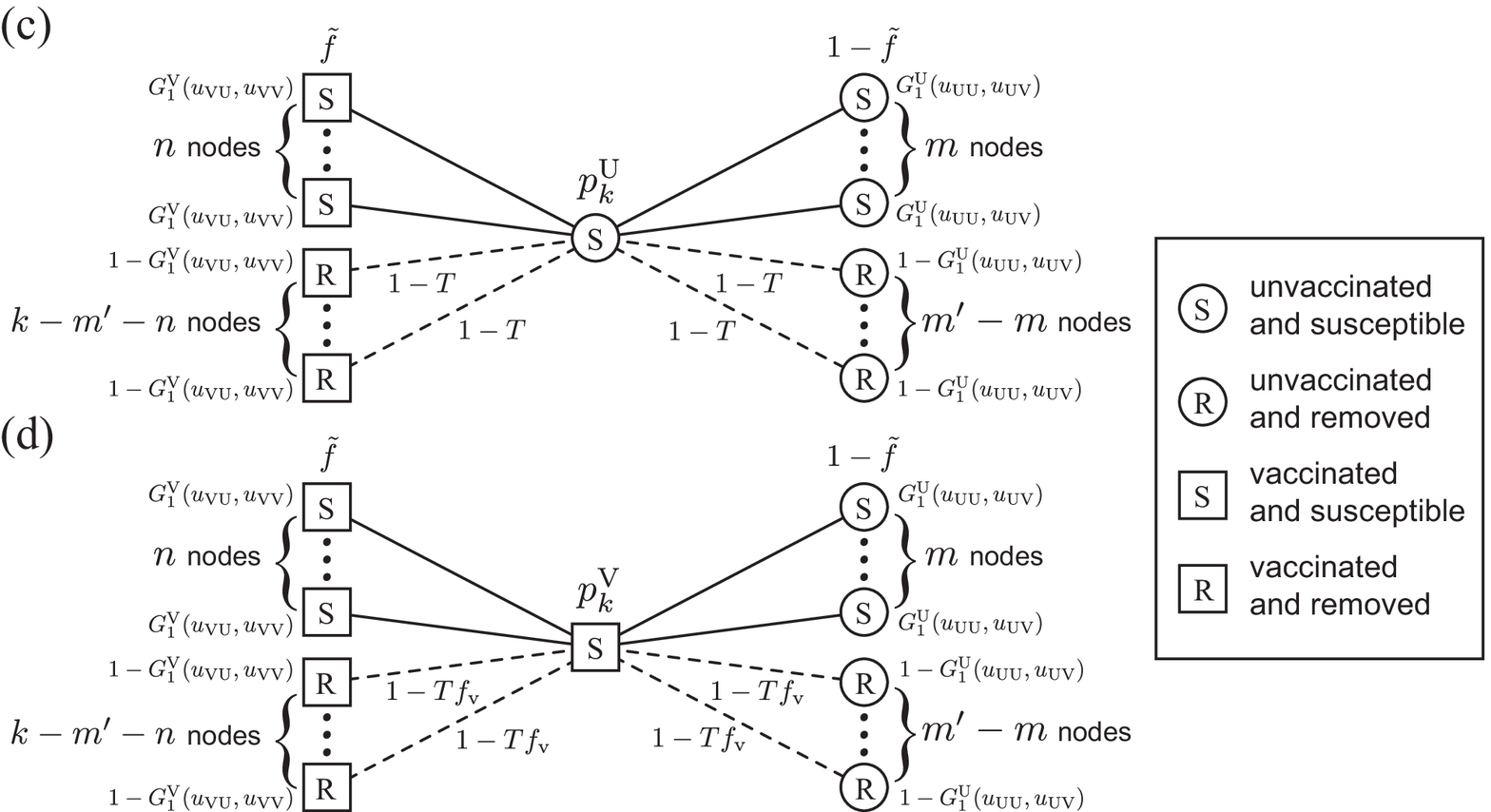}
\end{center}
\caption{
Schematics of (a) $F_0^{\rm U}(x)$, 
(b) $G_0^{\rm U}(x, y)$, 
(c) $P_0^{\rm U}(x, y)$, and (d) $P_0^{\rm V}(x, y)$.
}
\label{gf1}
\end{figure}


\subsection{SIR Model on Networks}

In the following, we consider uncorrelated, 
locally tree-like networks with arbitrary degree distributions.
In this section, we quickly review
the derivation of $T_{\rm c1}$ and $T_{\rm c2}$
in the absence of a defense strategy~\cite{Newman-Threshold-2005PRL,Newman-Spread-2002PRE}.


We denote by $p_k$ the fraction of nodes with degree $k$.
We denote by $q_k$ the distribution of the excess degree
defined as the degree of the node reached 
by following a randomly selected link minus 1;
$q_k=(k+1)p_{k+1}/\langle k \rangle$, 
where $\langle \cdot \rangle$ represents the average of a quantity weighted by $p_k$.
We define
\begin{eqnarray}
F_0(x) &\equiv \sum_k p_k x^k, \\
F_1(x) &\equiv \sum_k q_k x^k = \sum_k \frac{k p_k}{\langle k \rangle} x^{k-1}.
\end{eqnarray}
The giant component of the infected network is present if~\cite{Newman-Threshold-2005PRL}
\begin{equation}
T>T_{\rm c1}=\frac{\langle k \rangle}{\langle k^2-k \rangle}. \label{1stT}
\end{equation}


To derive $T_{\rm c2}$,
let $u$ be the probability that a node is not infected by a specific neighbor 
in a propagating attack.
After a propagating attack,
a node is 
susceptible with probability $\sum_k p_k u^k=F_0(u)$, 
and a node reached by following a link is susceptible with probability $\sum_k q_k u^k=F_1(u)$.
The following recursive relationship determines $u$:
\begin{equation}
u=1-T+T F_1(u). \label{urec}
\end{equation}
The probability that a node is susceptible in the residual network
and its randomly selected neighbor is
removed is equal to
$(1-T)[1-F_1(u)]=u-F_1(u)$.
Therefore, the probability that a node in the residual network 
has $m$ susceptible neighbors, i.e.,
the degree distribution of the residual network, is
represented as
\begin{equation}
\frac{1}{F_0(u)} \sum_{k=m}^\infty p_k 
{k \choose m}
F_1(u)^m [u-F_1(u)]^{k-m}. \nonumber
\end{equation}
The corresponding generating function $P_0(x)$ is given by
\begin{eqnarray}
P_0(x) 
&=
\frac{1}{F_0(u)} \sum_{m=0}^\infty \sum_{k=m}^\infty p_k 
{k \choose m}
F_1(u)^m [u-F_1(u)]^{k-m} x^m \nonumber \\
&=
\frac{1}{F_0(u)} \sum_{k=0}^\infty p_k [F_1(u) (x-1) +u]^k 
=\frac{F_0 [F_1(u) (x-1) +u]}{F_0(u)}.
\end{eqnarray}
Similarly, the generating function $P_1(x)$ 
for the excess degree distribution of the residual network is represented as
\begin{equation}
P_1(x) = \frac{F_1 [F_1(u) (x-1) +u]}{F_1(u)}. \label{P_1}
\end{equation}
From Eq.~(\ref{P_1}), the mean excess degree of the residual network, 
denoted by $\langle k_{\rm ex} \rangle_{\rm res}$, is calculated as
\begin{equation}
\langle k_{\rm ex} \rangle_{\rm res} =\frac{\rmd}{\rmd x} P_1(x) \Big|_{x=1}=F_1^{\prime}(u). 
\label{k_ex}
\end{equation}
The giant component of the residual network is absent if $T>T_{\rm c2}$, 
where $T_{\rm c2}$ is determined by
$\langle k_{\rm ex} \rangle_{\rm res}=F_1'(u)=1$~\cite{Newman-Threshold-2005PRL}.

\subsection{Defense strategies against propagating attacks \label{SecDef}}

In the following, we calculate $T_{\rm c1}$ and $T_{\rm c2}$
in the presence of
the random 
or degree-based defense.

\subsubsection{Degree Distributions of Vaccinated and Unvaccinated Nodes in the Original Network}

Under a defense strategy, each node is either
unvaccinated (U) or vaccinated (V) (figure~\ref{example}(b)).
First, we derive
the degree distribution 
of U nodes and that of V nodes in the original network.
Under both defense strategies,
a randomly selected node is U with probability $1-f$ and V otherwise.
We denote
the probability that a node reached by following a randomly selected link is 
a V node by $\tilde{f}$. Then, the probability that a node reached by following a
randomly selected link is a U node is equal to $1-\tilde{f}$.
Under the random defense, we obtain
\begin{equation}
\tilde{f}=f.
\end{equation}
Under the degree-based defense, we obtain
\begin{equation}
\tilde{f}= \sum_{k=k_{\rm cut}}^\infty q_k, 
\end{equation}
where $k_{\rm cut}$ is the minimum degree of V nodes 
determined by
\begin{equation}
\sum_{k=k_{\rm cut}}^\infty p_k =f.
\end{equation}

Let $p_k^{\rm U}$ and $p_k^{\rm V}$ be 
the degree distributions of U nodes and  V nodes, 
respectively, 
and $q_k^{\rm U}$ and $q_k^{\rm V}$ be 
the distributions of the excess degrees of U nodes and V nodes, respectively.
For the random defense, 
we obtain
\begin{eqnarray}
\{p_k^{\rm U}, p_k^{\rm V} \} &=\{ p_k, p_k \}, \label{pkrandom} \\ 
\{q_k^{\rm U}, q_k^{\rm V} \} &=\{ q_k, q_k \}, 
\end{eqnarray}
for all $k$.
For the degree-based defense, we obtain 
\begin{eqnarray}
\{p_k^{\rm U}, p_k^{\rm V} \}=
{\Biggl\{}
\begin{array}{ccl}
\{ p_k/(1-f), 0 \} & \qquad (k < k_{\rm cut}), & \\
\{ 0, p_k/f \} & \qquad (k_{\rm cut} \le k), & \label{pktarget}
\end{array}
\end{eqnarray}
\begin{eqnarray}
\{ q_k^{\rm U}, q_k^{\rm V} \}=
{\Biggl\{}
\begin{array}{ccl}
\{ q_k/(1-\tilde{f}), 0 \} & \qquad (k < k_{\rm cut}), & \\
\{ 0, q_k/\tilde{f} \} & \qquad (k_{\rm cut} \le k). & 
\end{array}
\end{eqnarray}

\subsubsection{Derivation of $T_{\rm c1}$}

We analyze the interaction of the generating functions for U nodes and those for V nodes. 
Related approaches have been adopted to analyze the percolation in networks that contain multiple types of nodes \cite{Soderberg2002,Leicht2009}.
We define the following generating functions 
\begin{eqnarray}
F_0^{\rm U}(x) &\equiv \sum_{k=0}^{\infty} p_k^{\rm U} x^k, \\ 
F_0^{\rm V}(x) &\equiv \sum_{k=0}^\infty p_k^{\rm V} x^k, \\
F_1^{\rm U}(x) &\equiv \sum_{k=0}^{\infty} q_k^{\rm U} x^k, \\
F_1^{\rm V}(x) &\equiv \sum_{k=0}^\infty q_k^{\rm V} x^k.
\end{eqnarray}
A schematic of $F_0^{\rm U}(x)$ is shown in figure~\ref{gf1}(a).

We extend the theory developed in~\cite{newman_random_2002}
to derive $T_{\rm c1}$ for our model. Consider the mean number
$N_{\ell}$ of removed nodes at distance $\ell$ from the
seed node.  We obtain
\begin{eqnarray}
\fl N_{\rm 1} &= \sum_{k=0}^\infty p_k \sum_{m=0}^k  
{k \choose m}
[m T +(k-m)T f_{\rm v}] (1-\tilde{f})^m \tilde{f}^{k-m} \nonumber \\
\fl &=
\Big( T \frac{\partial}{\partial x}+ T f_{\rm v} \frac{\partial}{\partial y} \Big) 
F_0[(1-\tilde{f})x+\tilde{f}y] \Big|_{x,y=1} 
=
[T (1-\tilde{f})+ T f_{\rm v} \tilde{f}] F_0'(1), 
\end{eqnarray}
\begin{eqnarray}
\fl N_{\rm 2} &= \sum_{k=0}^\infty  p_k \sum_{m=0}^k 
{k \choose m}
(1-\tilde{f})^m \tilde{f}^{k-m}
\nonumber \\
\fl &\quad \times \Big\{
m T \sum_{k'=0}^\infty q_{k'}^{\rm U} \sum_{m'=0}^{k'} 
{k' \choose m'}
[m' T +(k'-m') T f_{\rm v}] (1-\tilde{f})^{m'} \tilde{f}^{k'-m'} \nonumber \\
\fl &\quad\quad
+(k-m) T f_{\rm v} \sum_{k'=0}^\infty q_{k'}^{\rm V} \sum_{m'=0}^{k'} 
{k' \choose m'}
[m' T +(k'-m') T f_{\rm v}] (1-\tilde{f})^{m'} \tilde{f}^{k'-m'} \Big\} \nonumber \\
\fl 
&=
F_0'(1) [T (1-\tilde{f})+ T f_{\rm v} \tilde{f}]
[T (1-\tilde{f})  {F_1^{\rm U}}^{\prime}(1)+T f_{\rm v} \tilde{f} {F_1^{\rm V}}^{\prime}(1)], 
\end{eqnarray}
and 
\begin{eqnarray}
\fl N_{\ell} 
=
F_0'(1) [T (1-\tilde{f})+ T f_{\rm v} \tilde{f}]
[T (1-\tilde{f})  {F_1^{\rm U}}^{\prime}(1)+T f_{\rm v} \tilde{f} {F_1^{\rm V}}^{\prime}(1)]^{\ell-1} 
= \Big(\frac{N_{2} }{N_{1} } \Big)^{\ell-1} N_{1},
\end{eqnarray}
for any $\ell\ge 1$.
The infected network is a giant component if and only if
$N_{2} > N_{1}$, or equivalently, 
\begin{eqnarray}
\frac{N_{2}}{N_{1}} = 
T (1-\tilde{f}) {F_1^{\rm U}}^{\prime}(1)+T f_{\rm v} \tilde{f} {F_1^{\rm V}}^{\prime}(1)>1. \label{1stTcondition}
\end{eqnarray}
Therefore, $T_{\rm c1}$ is determined from 
\begin{eqnarray}
T_{\rm c1} 
[(1-\tilde{f}) {F_1^{\rm U}}^{\prime}(1)+f_{\rm v} \tilde{f} {F_1^{\rm V}}^{\prime}(1)]=1. \label{1stTdef}
\end{eqnarray}

In the case of the random defense, 
where ${F_1^{\rm U}}^{\prime}(1)={F_1^{\rm V}}^{\prime}(1)=F_1^{\prime}(1)$ and 
$\tilde{f}=f$,
Eq.~(\ref{1stTdef}) is reduced to 
\begin{eqnarray}
 T_{\rm c1} [(1-f)+ f_{\rm v} f] \frac{\langle k(k-1) \rangle}{\langle k \rangle} =1. \label{1stTdefRand}
\end{eqnarray}

\subsubsection{Derivation of $T_{\rm c2}$}

The probability that a randomly selected U node has exactly $m$ U neighbors 
is equal to
$\sum_{k=m}^\infty p_k^{\rm U} 
{k \choose m}
(1-\tilde{f})^m \tilde{f}^{k-m}$.
The generating function $G_0^{\rm U}(x,y)$ for this probability distribution, which is
schematically shown in figure~\ref{gf1}(b), is given by 
\begin{equation}
G_0^{\rm U}(x,y)\equiv
\sum_ {k=0}^\infty p_k^{\rm U} \sum_{m=0}^{k} 
 {k \choose m}
(1-\tilde{f})^m \tilde{f}^{k-m} x^m y^{k-m} =
F_0^{\rm U}[(1-\tilde{f})x+\tilde{f}y].
\end{equation}
The generating function $G_0^{\rm V}(x,y)$ for the probability distribution that 
a $V$ node has $m$ U neighbors 
is given by
\begin{equation}
G_0^{\rm V}(x,y)\equiv
\sum_ {k=0}^\infty p_k^{\rm V} \sum_{m=0}^{k} 
 {k \choose m}
(1-\tilde{f})^m \tilde{f}^{k-m} x^m y^{k-m} =
F_0^{\rm V}[(1-\tilde{f})x+\tilde{f}y].
\end{equation}
Similarly, we obtain
\begin{equation}
G_1^{\rm U}(x,y)\equiv
\sum_ {k=0}^\infty q_k^{\rm U} \sum_{m=0}^{k} 
 {k \choose m}
(1-\tilde{f})^m \tilde{f}^{k-m} x^m y^{k-m} 
= F_1^{\rm U}[(1-\tilde{f})x+\tilde{f}y], 
\end{equation}
\begin{equation}
G_1^{\rm V}(x,y)\equiv\sum_ {k=0}^\infty q_k^{\rm V} \sum_{m=0}^{k} 
 {k \choose m}
(1-\tilde{f})^m \tilde{f}^{k-m} x^m y^{k-m} =
F_1^{\rm V}[(1-\tilde{f})x+\tilde{f}y].
\end{equation}

Let $u_{\rm XY}$ ($X,Y \in \{ U,V \}$) be the probability that 
an X node is not infected by a specific Y neighbor in a propagating attack.
Then, a U node is susceptible with 
probability $G_0^{\rm U}(u_{\rm UU},u_{\rm UV})$, and 
a V node is susceptible with probability $G_0^{\rm V}(u_{\rm VU},u_{\rm VV})$.
A U node connected to
a randomly selected link is susceptible with probability 
$G_1^{\rm U}(u_{\rm UU},u_{\rm UV})$, 
and a V node connected to a random selected link is susceptible with probability $G_1^{\rm V}(u_{\rm VU},u_{\rm VV})$.
Then, 
$u_{\rm UU}$, $u_{\rm UV}$, $u_{\rm VU}$, and $u_{\rm VV}$ satisfy 
the following recursive relationships:
\begin{eqnarray}
u_{\rm UU} &= 1-T+T G_1^{\rm U}(u_{\rm UU},u_{\rm UV}), \label{Uuu} \\
u_{\rm UV} &= 1-T+T G_1^{\rm V}(u_{\rm VU},u_{\rm VV}),  \label{Uuv} \\
u_{\rm VU} &= 1-T f_{\rm v}+T f_{\rm v} G_1^{\rm U}(u_{\rm UU},u_{\rm UV}),  \label{Uvu} \\
u_{\rm VV} &= 1-T f_{\rm v}+T f_{\rm v} G_1^{\rm V}(u_{\rm VU},u_{\rm VV}). \label{Uvv}
\end{eqnarray}

To derive the degree distribution of U nodes
in the residual network, consider a focal U node.
When the focal U node is adjacent to a U node, 
the probability that the U neighbor is (infected and) eventually removed 
and the U focal node is not infected by this U neighbor in a propagating attack 
is equal to 
\begin{equation}
(1-T)[1-G_1^{\rm U}(u_{\rm UU},u_{\rm UV})]
=u_{\rm UU}-G_1^{\rm U}(u_{\rm UU},u_{\rm UV}).
 \nonumber
\end{equation}
When the focal U node is adjacent to a V node, 
the probability that the V neighbor is eventually removed 
and the focal U node is not infected by this V neighbor
is equal to 
\begin{equation}
(1-T)[1-G_1^{\rm V}(u_{\rm VU},u_{\rm VV})]
=u_{\rm UV}-G_1^{\rm V}(u_{\rm VU},u_{\rm VV}).
 \nonumber
\end{equation}
Then, the probability that a U node has $m$ susceptible U neighbors and 
$n$ susceptible V neighbors 
and is not infected 
is equal to
\begin{eqnarray}
\fl & \sum_{k=m+n}^\infty \sum_{m'=m}^{k-n} p_k^{\rm U} 
 {k \choose m'}
 (1-\tilde{f})^{m'}\tilde{f}^{k-m'} 
 {m' \choose m}
G_1^{\rm U}(u_{\rm UU},u_{\rm UV})^m 
[u_{\rm UU}-G_1^{\rm U}(u_{\rm UU},u_{\rm UV})]^{m'-m} \nonumber \\
\fl &\times 
 {k-m' \choose n}
G_1^{\rm V}(u_{\rm VU},u_{\rm VV})^n 
[u_{\rm UV}-G_1^{\rm V}(u_{\rm VU},u_{\rm VV})]^{k-m'-n}. \nonumber
\end{eqnarray}
Therefore, the generating function $P_0^{\rm U}(x,y)$ 
for the degree distribution of U nodes in the residual network 
(figure~\ref{gf1}(c)) is given by 
\begin{eqnarray}
\fl P_0^{\rm U}(x,y)&=
\frac{1}{G_0^{\rm U}(u_{\rm UU}, u_{\rm UV})}
\sum_{k=0}^\infty p_k^{\rm U} \sum_{m'=0}^{k}  
  {k \choose m'}
(1-\tilde{f})^{m'}\tilde{f}^{k-m'} \nonumber \\
\fl & \times 
\sum_{m=0}^{m'} 
 {m' \choose m}
G_1^{\rm U}(u_{\rm UU},u_{\rm UV})^m 
[u_{\rm UU}-G_1^{\rm U}(u_{\rm UU},u_{\rm UV})]^{m'-m} x^m \nonumber \\
\fl & \times \sum_{n=0}^{k-m'} 
 {k-m' \choose n}
G_1^{\rm V}(u_{\rm VU},u_{\rm VV})^n 
[u_{\rm UV}-G_1^{\rm V}(u_{\rm VU},u_{\rm VV})]^{k-m'-n} y^n \nonumber \\
\fl &= 
\frac{G_0^{\rm U}[G_1^{\rm U}(u_{\rm UU},u_{\rm UV})(x-1)+u_{\rm UU}, 
G_1^{\rm V}(u_{\rm VU},u_{\rm VV})(y-1)+u_{\rm UV}]}
{G_0^{\rm U}(u_{\rm UU}, u_{\rm UV})}.
\end{eqnarray}

We can similarly derive the degree distribution of V nodes in the residual network. 
When a focal V node is adjacent to a U node, 
the probability that the U neighbor is eventually removed 
and the focal V node is not infected by this U neighbor
is equal to 
\begin{equation}
(1-T f_{\rm v})[1-G_1^{\rm U}(u_{\rm UU},u_{\rm UV})]
=u_{\rm VU}-G_1^{\rm U}(u_{\rm UU},u_{\rm UV}).
\nonumber
\end{equation}
When the focal V node is adjacent to a V node, 
the probability that the V neighbor is eventually removed 
and the focal V node is not infected by this V neighbor is equal to 
\begin{equation}
(1-T f_{\rm v})[1-G_1^{\rm V}(u_{\rm VU},u_{\rm VV})]
=u_{\rm VV}-G_1^{\rm V}(u_{\rm VU},u_{\rm VV}). \nonumber
\end{equation}
The generating function $P_0^{\rm V}(x,y)$ 
for the degree distribution of V nodes in the residual network 
(figure~\ref{gf1}(d)) is given by
\begin{eqnarray}
\fl P_0^{\rm V}(x,y)&=
\frac{1}{G_0^{\rm V}(u_{\rm VU}, u_{\rm VV})} 
\sum_{k=0}^\infty p_k^{\rm V} \sum_{m'=0}^{k}  
{k \choose m'}
(1-\tilde{f})^{m'}\tilde{f}^{k-m'} \nonumber \\
\fl & \times \sum_{m=0}^{m'} 
{m' \choose m}
G_1^{\rm U}(u_{\rm UU},u_{\rm UV})^m 
[u_{\rm VU}-G_1^{\rm U}(u_{\rm UU},u_{\rm UV})]^{m'-m} x^m \nonumber \\
\fl & \times \sum_{n=0}^{k-m'} 
{k-m' \choose n}
G_1^{\rm V}(u_{\rm VU},u_{\rm VV})^n 
[u_{\rm VV}-G_1^{\rm V}(u_{\rm VU},u_{\rm VV})]^{k-m'-n} y^n
 \nonumber \\
\fl &= 
\frac{
G_0^{\rm V}[G_1^{\rm U}(u_{\rm UU},u_{\rm UV})(x-1)+u_{\rm VU}, 
G_1^{\rm V}(u_{\rm VU},u_{\rm VV})(y-1)+u_{\rm VV}]
}{G_0^{\rm V}(u_{\rm VU}, u_{\rm VV})}.
\end{eqnarray}

Similarly, the generating functions for the excess degree distributions 
of U nodes and V nodes in the residual network are given by 
\begin{eqnarray}
\fl P_1^{\rm U}(x,y)=
\frac{
G_1^{\rm U}[G_1^{\rm U}(u_{\rm UU},u_{\rm UV})(x-1)+u_{\rm UU}, 
G_1^{\rm V}(u_{\rm VU},u_{\rm VV})(y-1)+u_{\rm UV}]
}
{G_1^{\rm U}(u_{\rm UU}, u_{\rm UV})},
\end{eqnarray}
and 
\begin{eqnarray}
\fl P_1^{\rm V}(x,y)=
\frac{
G_1^{\rm V}[G_1^{\rm U}(u_{\rm UU},u_{\rm UV})(x-1)+u_{\rm VU}, 
G_1^{\rm V}(u_{\rm VU},u_{\rm VV})(y-1)+u_{\rm VV}]
}
{G_1^{\rm V}(u_{\rm VU}, u_{\rm VV})},
\end{eqnarray}
respectively.

When we follow a randomly selected link in the residual network, 
we reach a U node with probability 
\begin{eqnarray}
\frac{(1-\tilde{f}) G_1^{\rm U}(u_{\rm UU}, u_{\rm UV})}{
(1-\tilde{f}) G_1^{\rm U}(u_{\rm UU}, u_{\rm UV})+ \tilde{f} G_1^{\rm V}(u_{\rm VU}, u_{\rm VV})}. \nonumber
\end{eqnarray}
Otherwise, we reach a V node.
Then, the mean excess degree of the residual network is given by 
\begin{eqnarray}
\fl 
\langle k_{\rm ex} \rangle_{\rm res} 
&=
\Big(\frac{\partial}{\partial x}+\frac{\partial}{\partial y} \Big) 
\frac{(1-\tilde{f}) G_1^{\rm U}(u_{\rm UU}, u_{\rm UV}) P_1^{\rm U}(x,y)+\tilde{f} G_1^{\rm V}(u_{\rm VU}, u_{\rm VV}) P_1^{\rm V}(x,y)}
{(1-\tilde{f}) G_1^{\rm U}(u_{\rm UU}, u_{\rm UV})+ \tilde{f} G_1^{\rm V}(u_{\rm VU}, u_{\rm VV})}
\Big|_{x,y=1} \nonumber \\
\fl &= 
(1-\tilde{f}) {F_1^{\rm U}}^{\prime}[ (1-\tilde{f}) u_{\rm UU}+ \tilde{f} u_{\rm UV}]
+\tilde{f}{F_1^{\rm V}}^{\prime}[ (1-\tilde{f}) u_{\rm VU} +\tilde{f} u_{\rm VV}].
\end{eqnarray}
Because the second transition point is given by 
$\langle k_{\rm ex} \rangle_{\rm res}=1$, 
we find $T_{\rm c2}$ by solving
\begin{eqnarray}
(1-\tilde{f}) {F_1^{\rm U}}^{\prime}[(1-\tilde{f}) u_{\rm UU}+\tilde{f} u_{\rm UV}] 
+\tilde{f} {F_1^{\rm V}}^{\prime}[(1-\tilde{f}) u_{\rm VU}+\tilde{f} u_{\rm VV}] =1. \label{2ndTdef}
\end{eqnarray}

\subsubsection{Component Size}

The generating function formalism also gives
the component size of the infected and residual networks.
Because a randomly selected U node is susceptible 
with probability $G_0^{\rm U}(u_{\rm UU}, u_{\rm UV})$ and 
a randomly selected V node is susceptible with probability $G_0^{\rm V}(u_{\rm VU}, u_{\rm VV})$, 
the component size of the infected network $S_{\rm inf}$ is represented as
\begin{equation}
S_{\rm inf}=1-(1-f) G_0^{\rm U}(u_{\rm UU}, u_{\rm UV}) -f G_0^{\rm V}(u_{\rm VU}, u_{\rm VV}). \label{sizeS}
\end{equation}

To derive the largest component size of the residual network, 
we denote by $H_0^{\rm U}(x)$ and $H_0^{\rm V}(x)$ the generating functions 
for the size of the finite components of the residual networks 
to which a randomly selected U node and V node belong~\cite{newman2003structure}.
The average size $S_{\rm res}$ of the 
largest component of the residual network relative to 
the size of the original network is given by 
\begin{equation}
S_{\rm res}=(1-S_{\rm inf})\left[1-
\frac{G_0^{\rm U}(u_{\rm UU}, u_{\rm UV}) H_0^{\rm U}(1)+G_0^{\rm V}(u_{\rm VU}, u_{\rm VV})H_0^{\rm V}(1)}
{G_0^{\rm U}(u_{\rm UU}, u_{\rm UV})+G_0^{\rm V}(u_{\rm VU}, u_{\rm VV})}\right],
\label{sizeC}
\end{equation}
where
\begin{eqnarray}
H_0^{\rm U}(1) &= P_0^{\rm U}(A,B), \\ 
H_0^{\rm V}(1) &= P_0^{\rm V}(A,B),
\end{eqnarray}
\begin{eqnarray}
A &\equiv H_1^{\rm U}(1) = P_1^{\rm U}(A,B), \\
B &\equiv H_1^{\rm V}(1) = P_1^{\rm V}(A,B).
\end{eqnarray}

\subsubsection{Degree-dependent Probability of Infection}

We denote by $U(k)$ the probability that 
a node with degree $k$ is uninfected.
Because 
a U node is susceptible with probability $G_0^{\rm U}(u_{\rm UU},u_{\rm UV})$ 
and a V node is susceptible with probability $G_0^{\rm V}(u_{\rm VU},u_{\rm VV})$, 
$U(k)$ is given by 
\begin{eqnarray}
\fl
\sum_{k=0}^{\infty} p_k U(k) 
&=
(1-f) G_0^{\rm U}(u_{\rm UU},u_{\rm UV}) + f G_0^{\rm V}(u_{\rm VU},u_{\rm VV}) \nonumber \\
\fl
&=
(1-f) \sum_{k=0}^{\infty} p_k^{\rm U} [(1-\tilde{f})u_{\rm UU}+\tilde{f} u_{\rm UV}]^k
+
f \sum_{k=0}^{\infty} p_k^{\rm V} [(1-\tilde{f})u_{\rm VU}+\tilde{f} 
u_{\rm VV}]^k. \label{Uk}
\end{eqnarray}
For the random defense, 
$p_k^{\rm U}$ and $p_k^{\rm V}$ are given by Eq.~(\ref{pkrandom}). Therefore,
Eq.~(\ref{Uk}) is reduced to 
\begin{eqnarray}
\fl
\sum_{k=0}^{\infty} p_k U(k) 
=
\sum_{k=0}^{\infty} p_k 
\{ (1-f) [(1-f)u_{\rm UU}+f u_{\rm UV}]^k
+
f [(1-f)u_{\rm VU}+f 
u_{\rm VV}]^k \}. \label{Ukrandom} 
\end{eqnarray}
Because Eq.~(\ref{Ukrandom}) is satisfied for an arbitrary degree
distribution $\{p_k\}$, 
we obtain
\begin{equation}
U(k) = (1-f) [(1-f) u_{\rm UU}+f u_{\rm UV}]^k
+ f [(1-f) u_{\rm VU}+f u_{\rm VV}]^k.
\label{resultUkrandom} 
\end{equation}
For the degree-based defense, 
$p_k^{\rm U}$ and $p_k^{\rm V}$ are given by Eq.~(\ref{pktarget}).
Therefore, Eq.~(\ref{Uk}) is reduced to 
\begin{eqnarray}
\fl
\sum_{k=0}^{\infty} p_k U(k) =
\sum_{k=0}^{k_{\rm cut}-1}
p_k [(1-\tilde{f})u_{\rm UU}+\tilde{f} u_{\rm UV}]^k 
+\sum_{k=k_{\rm cut}}^{\infty}
p_k [(1-\tilde{f})u_{\rm VU}+\tilde{f} u_{\rm VV}]^k. \label{Uktarget}
\end{eqnarray}
Because Eq.~(\ref{Uktarget}) is satisfied for arbitrary $\{p_k\}$, 
we obtain  
\begin{eqnarray}
U(k) = {\Biggl\{}
\begin{array}{ccl}
[(1-\tilde{f})u_{\rm UU}+\tilde{f} u_{\rm UV}]^k & \qquad (k < k_{\rm cut}), 
& \\

[(1-\tilde{f})u_{\rm VU}+\tilde{f} u_{\rm VV}]^k & \qquad (k_{\rm cut} \le k). & 
\end{array}
\label{resultUktarget}
\end{eqnarray}

\section{Numerical Results}


\begin{figure}
\begin{center}
\includegraphics[width=5.cm]{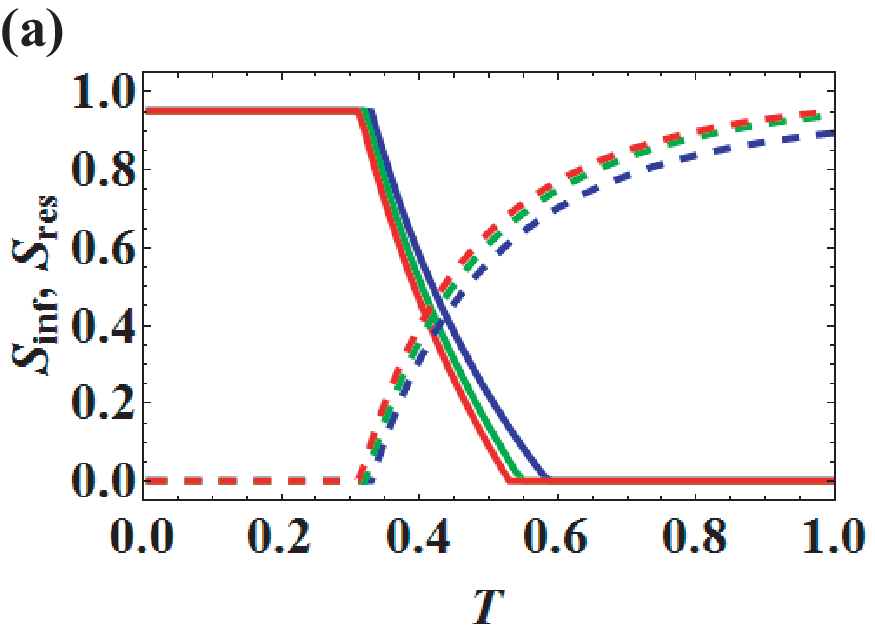}
\includegraphics[width=5.cm]{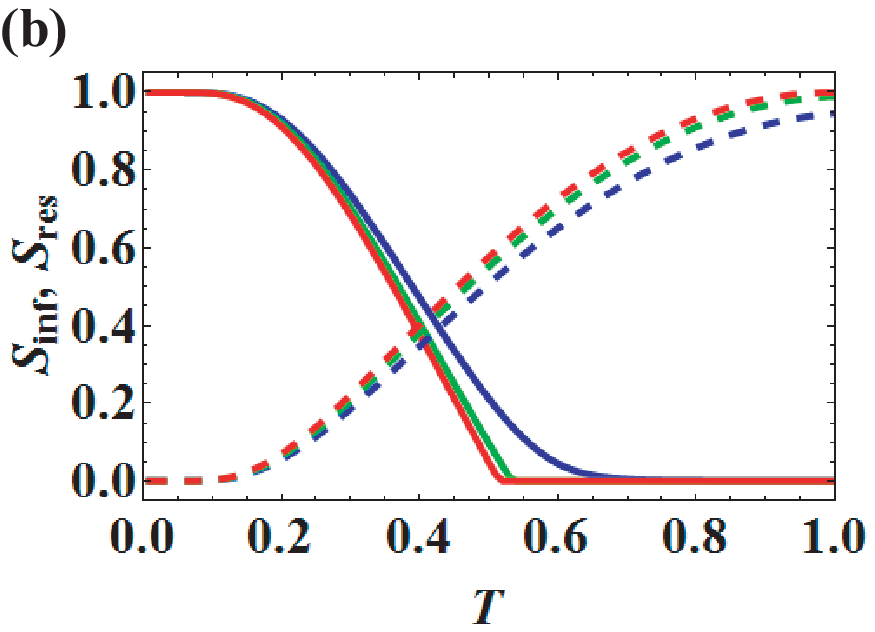}
\includegraphics[width=5.cm]{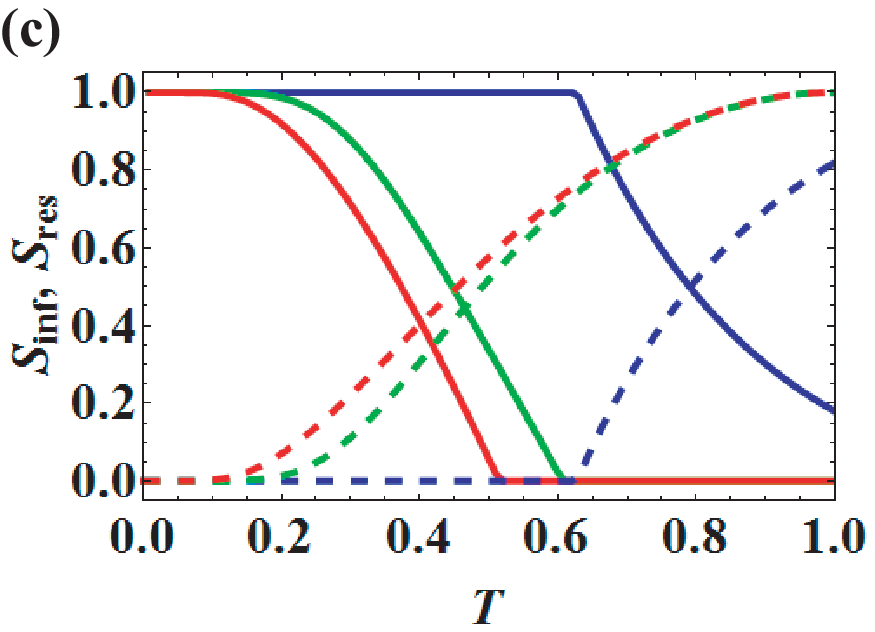}
\end{center}
\caption{
Component size $S_{\rm inf}$ of the infected network (dashed lines) and 
the largest component size $S_{\rm res}$ of the residual network (solid lines) 
for (a) random graph with the random defense (case (i)), (b) SF network with the random defense (case (ii)),
and (c) SF network with the degree-based defense (case (iii)).
We set $f=0.05$.
The different lines
correspond to $f_{\rm v}=0.00$, $0.50$, and $1.00$, from right to left.
}
\label{Comp-ER-rand}
\end{figure}


On the basis of the theoretical results obtained in Sec.~\ref{GFA}, we
numerically examine the robustnesses of different networks
against the propagating attack. 
We consider the following three
combinations of networks and defense strategies: 
\begin{description}
\item[(i)] random graph with random defense,
\item[(ii)] uncorrelated SF network with random defense, and
\item[(iii)] uncorrelated SF network with degree-based defense.
\end{description}
We assume that the SF network 
has degree distribution $p_k \propto k^{-3}$, 
where $k\ge k_{\rm min}\equiv 2$, which yields $\langle k \rangle \simeq 3.19$. 
For a fair comparison, 
we assume that $\langle k \rangle$ of the random graph is equal to
that of the SF network.
For the random graph, we obtain
$F_0(x)=F_1(x)=\exp [\langle k \rangle (x-1)]$.
For the SF network, we obtain 
$F_0(x) =\sum_{k=0}^\infty p_k x^k={\rm Li}_{3}(x)/{\rm Li}_{3}(1)$, 
$F_1(x) =\sum_{k=0}^\infty k p_k x^{k-1}/\sum_{k=0}^\infty k p_k={\rm Li}_{2}(x)/x {\rm Li}_{2}(1)$, 
where ${\rm Li}_{n}(x)$ is the $n$th polylogarithm of $x$, i.e.,
${\rm Li}_{n}(x)\equiv \sum_{k=1}^\infty x^k k^{-n}$.
In the random defense (cases (i) and (ii)), 
$F_0^{\rm U}(x)=F_0^{\rm V}(x)=F_0(x)$ and 
$F_1^{\rm U}(x)=F_1^{\rm V}(x)=F_1(x)$. 
In the degree-based defense (case (iii)), we vaccinate
all the nodes whose degree is equal to or larger than $k_{\rm cut}$ such that
the fraction of V nodes does not exceed $f$. Then, we vaccinate 
some nodes with degree $k_{\rm cut}-1$ such that the total fraction of V nodes
is equal to $f$. The other nodes with degree $k_{\rm cut}-1$ are U nodes.
The generating functions for U and V nodes are given by
\begin{eqnarray}
F_0^{\rm U}(x) &= 
\sum_{k=0}^{k_{\rm cut}^\prime} p_k x^k 
- C p_{k_{\rm cut}^\prime} x^{k_{\rm cut}^\prime}, \\ 
F_1^{\rm U}(x) &= 
\sum_{k=0}^{k_{\rm cut}^\prime} q_k x^k 
- C q_{k_{\rm cut}^\prime} x^{k_{\rm cut}^\prime}, \\ 
F_0^{\rm V}(x) &= 
\sum_{k=k_{\rm cut}}^{\infty} p_k x^k 
+ C p_{k_{\rm cut}^\prime} x^{k_{\rm cut}^\prime}, \\
F_1^{\rm V}(x) &= 
\sum_{k=k_{\rm cut}}^{\infty} q_k x^k 
+ C q_{k_{\rm cut}^\prime} x^{k_{\rm cut}^\prime},
\end{eqnarray}
where $C = (f- \sum_{k=k_{\rm cut}}^{\infty} p_k)/p_{k_{\rm cut}^\prime}$ and
$k_{\rm cut}^\prime =k_{\rm cut}-1$.
For each case, we apply the results of Sec.~\ref{SecDef} 
to obtain the component sizes and transition points. 
For given $f$ and $f_{\rm v}$, we substitute 
$F_0^{\rm U}(x)$, $F_1^{\rm U}(x)$, $F_0^{\rm V}(x)$, and $F_1^{\rm V}(x)$ 
in Eq.~(\ref{1stTdef}) to obtain $T_{\rm c1}$.
We also substitute 
$F_0^{\rm U}(x)$, $F_1^{\rm U}(x)$, $F_0^{\rm V}(x)$, and $F_1^{\rm V}(x)$ 
in Eqs.~(\ref{Uuu})--(\ref{Uvv}) to determine 
$u_{\rm UU}$, $u_{\rm UV}$, $u_{\rm VU}$, and $u_{\rm VV}$. 
Then, we use the obtained values of 
$u_{\rm UU}$, $u_{\rm UV}$, $u_{\rm VU}$, and $u_{\rm VV}$ to derive
$T_{\rm c2}$, $S_{\rm inf}$, and $S_{\rm ref}$ 
from Eqs.~(\ref{2ndTdef}), (\ref{sizeS}), and (\ref{sizeC}), respectively.

We first examine the component sizes $S_{\rm inf}$ of the infected network and
$S_{\rm res}$ of the residual network using Eqs.~(\ref{sizeS}) and (\ref{sizeC}). 
The results for case (i)
with $f=0.05$ are shown in figure~\ref{Comp-ER-rand}(a).  When
$T$ is small, $S_{\rm inf}=0$ (in the limit of the infinite network size) holds
true irrespective of the value of $f_{\rm v}$, as shown 
by different dashed lines in figure~\ref{Comp-ER-rand}(a).  
When $T>T_{\rm c1}\approx 0.33$, $S_{\rm inf}$ increases and $S_{\rm res}$ decreases with an
increase in $T$.  When $T>T_{\rm c2} \approx 0.6$, we obtain
$S_{\rm res}=0$.  The precise values of $T_{\rm c1}$ and $T_{\rm c2}$
vary with $f_{\rm v}$.  Efficient vaccines (i.e., small
$f_{\rm v}$) make $S_{\rm inf}$ small and $S_{\rm res}$ large and increase
$T_{\rm c1}$ and $T_{\rm c2}$.

The results for case (ii)
are shown in
figure~\ref{Comp-ER-rand}(b).
$T_{\rm c1}$ is much smaller than in case (i);
Eq.~(\ref{1stTdefRand}) implies $T_{\rm c1}=0$ when ${F_1^{\rm V}}^{\prime}(1)$ diverges.
For $f_{\rm v} \neq 0$,
$S_{\rm res}$ behaves qualitatively the same as
that for case (i).
In particular, the value of $T_{\rm c2}$ for $f_{\rm v}=1$
is only slightly smaller than that for case (i).
For $f_{\rm v}=0$,
$S_{\rm res}$ behaves differently; 
$S_{\rm res}>0$ even when $T=1$, 
implying that the residual network is never destroyed.

The results for case (iii)
are shown in 
figure~\ref{Comp-ER-rand}(c).
We find that the degree-based defense is effective at decreasing (increasing) $S_{\rm inf}$ ($S_{\rm res}$) 
as compared with the random defense (case (ii)).
In particular, the perfect vaccines (i.e., $f_{\rm v} = 0$)
contain the propagating attack 
when $T<T_{\rm c1} \approx 0.63$.
We find from Eq.~(\ref{1stTdef}) that 
$T_{\rm c1} \neq 0$ only for $f_{\rm v} = 0$.
Therefore,
any leaky vaccination (i.e., $f_{\rm v} >0$) fails to contain the attack 
because ${F_1^{\rm V}}^{\prime}(1)$ diverges.
We remark that, for $f_{\rm v}=0$, 
Eq.~(\ref{1stTdef}) is reduced to a known result~\cite{Madar2004EPJB}
\begin{equation}
T_{\rm c1} \sum_{k=0}^{k_{\rm cut}-1} \frac{k(k-1)}{\langle k \rangle} p_k=1. \label{Tc1target-perfect}
\end{equation}
We also remark that, for $f_{\rm v}=1$, the value of $T_{\rm c2}$ coincides
with that for case (ii) as expected; the effect of vaccination is completely
null when $f_{\rm v}=1$.


\begin{figure}
\begin{center}
\includegraphics[height=4.5cm]{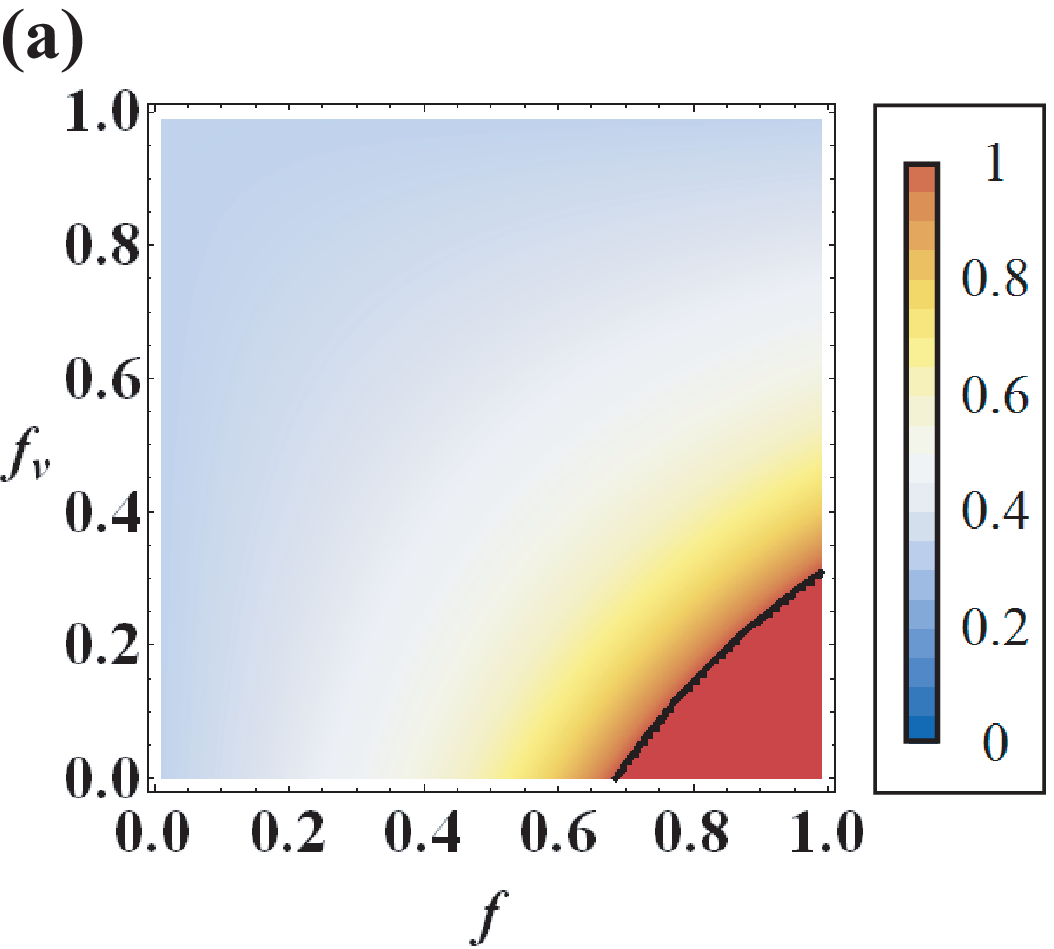}
\includegraphics[height=4.5cm]{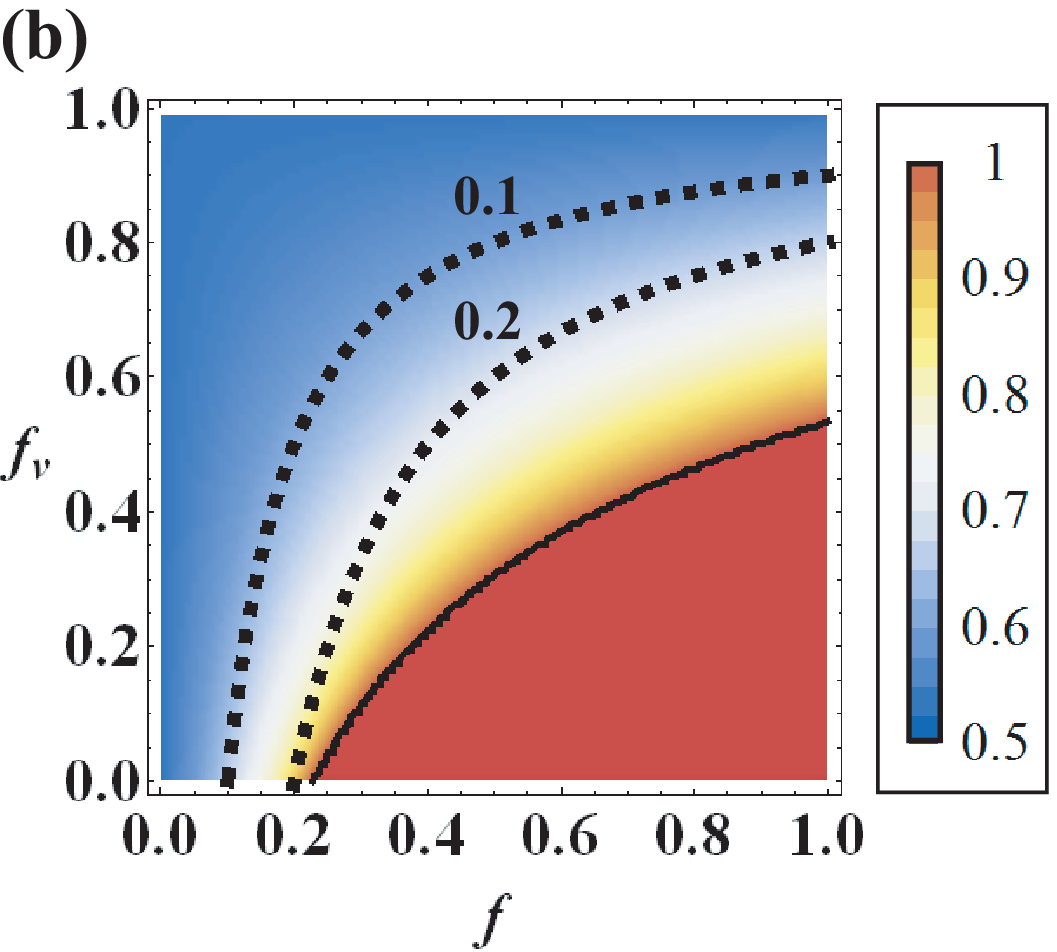}
\includegraphics[height=4.5cm]{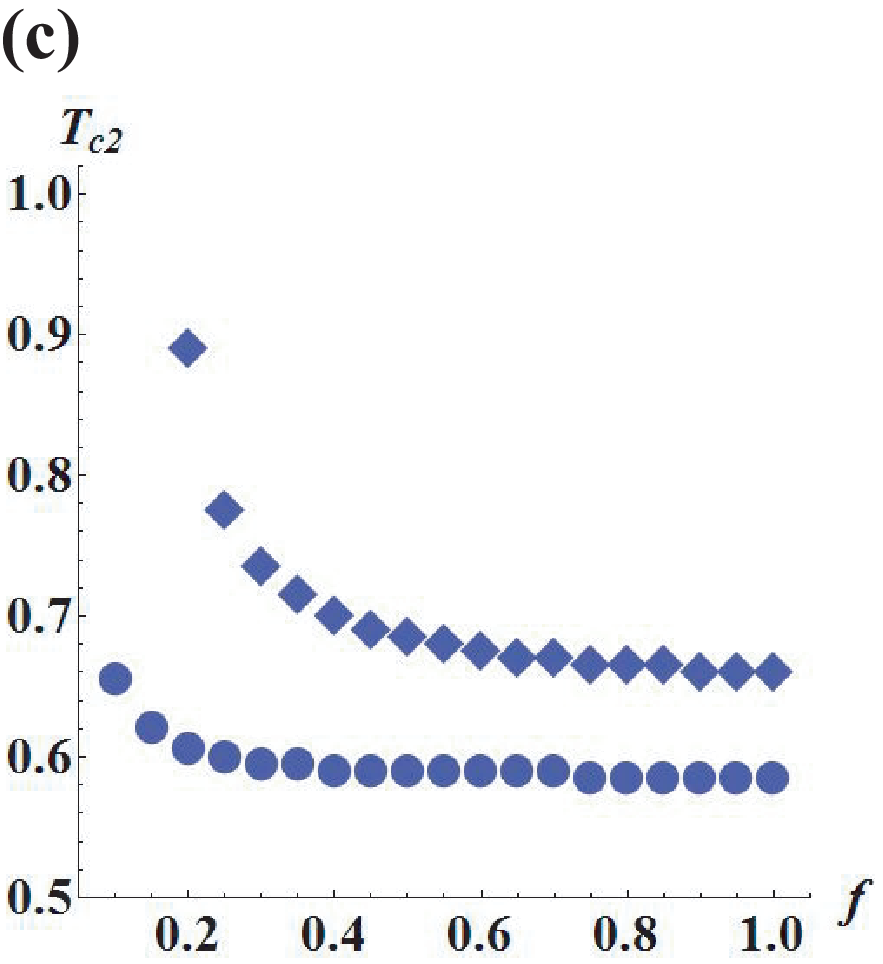}
\end{center}
\caption{Results for the random graph with the random defense (case (i)).
(a) $T_{\rm c1}$ and (b) $T_{\rm c2}$
as functions of $f$ and $f_{\rm v}$.
The solid lines indicate $T_{\rm c1}=1$ in (a) and $T_{\rm c2}=1$ in (b). 
The dashed lines indicate $f(1-f_{\rm v})=0.1$ and $f(1-f_{\rm v})=0.2$.
(c) $T_{\rm c2}$ conditioned by $f(1-f_{\rm v})=0.1$ (circles) and $0.2$ (diamonds).
}
\label{Phase-ER}
\end{figure}

\begin{figure}
\begin{center}
\includegraphics[height=5cm]{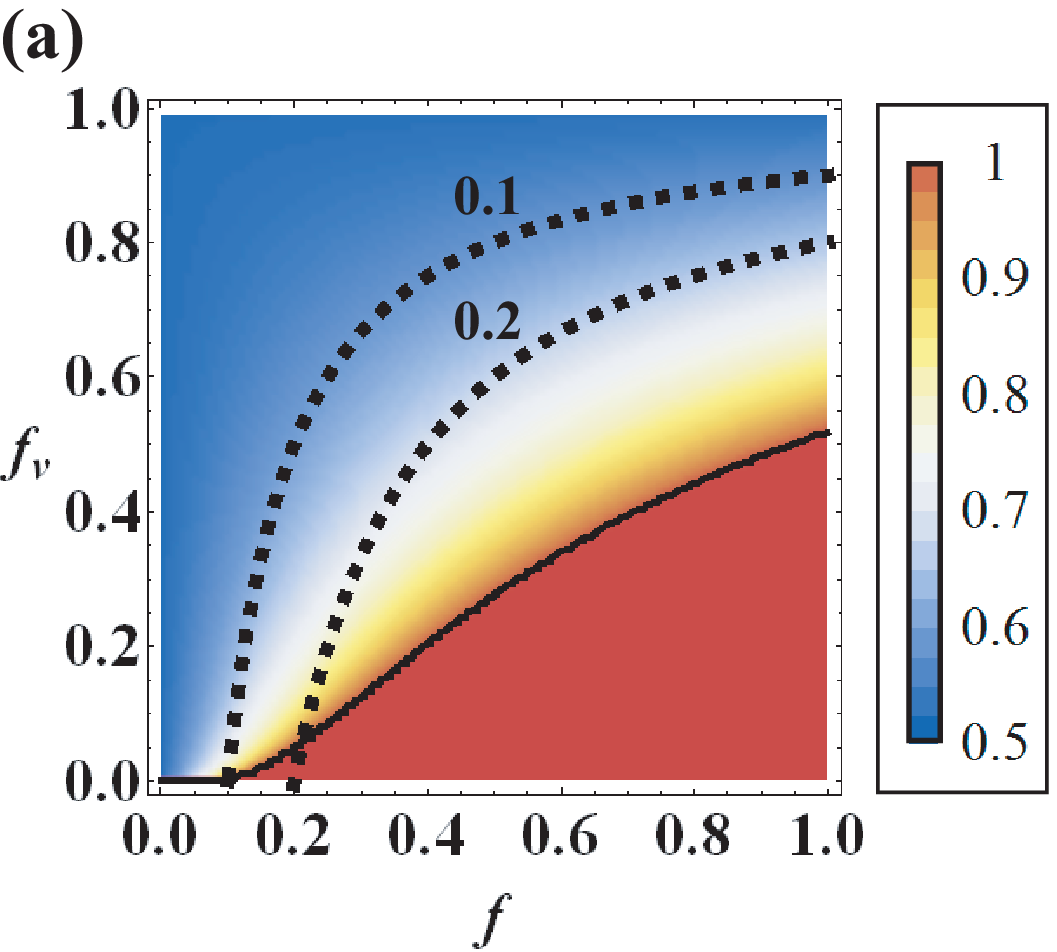}
\includegraphics[height=5cm]{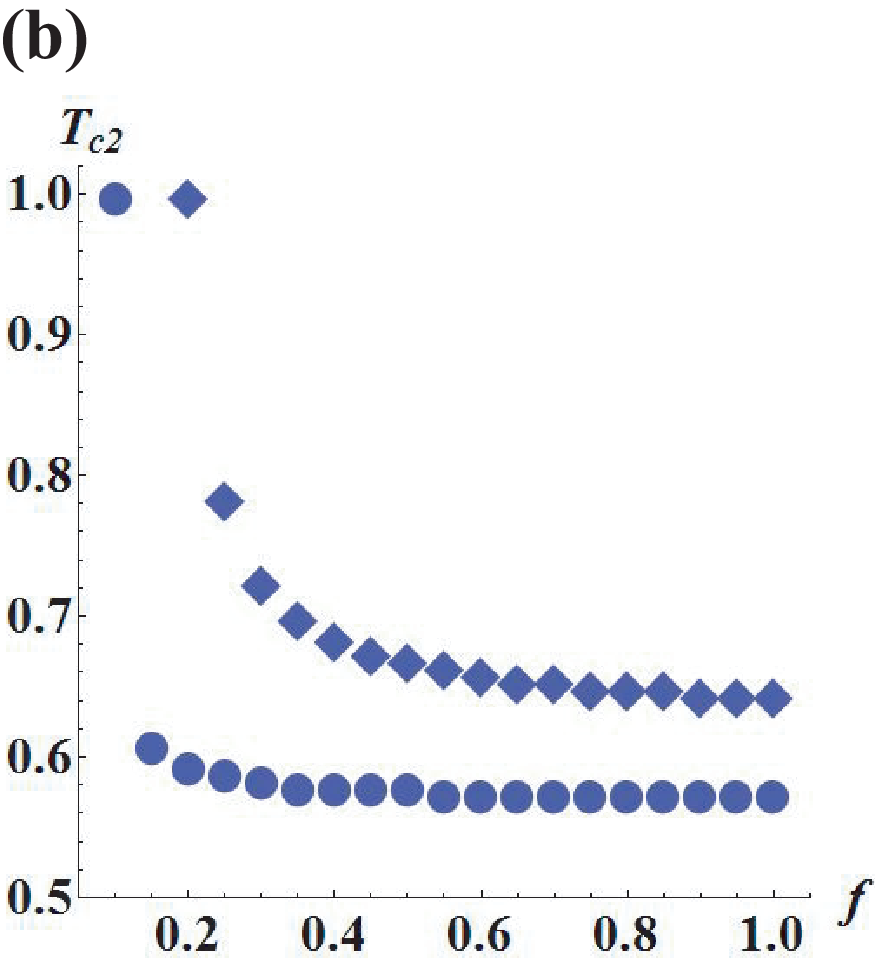}
\end{center}
\caption{Results for the SF network with random defense (case (ii)).
(a) $T_{\rm c2}$ as a function of $f$ and $f_{\rm v}$. 
The solid line indicates $T_{\rm c2}=1$.
The dashed lines indicate $f(1-f_{\rm v})=0.1$ and $f(1-f_{\rm v})=0.2$.
(b) $T_{\rm c2}$ conditioned by $f(1-f_{\rm v})=0.1$ (circles) and $0.2$ (diamonds).
}
\label{Phase-SFN-Rand}
\end{figure}

\begin{figure}
\begin{center}
\includegraphics[height=5cm]{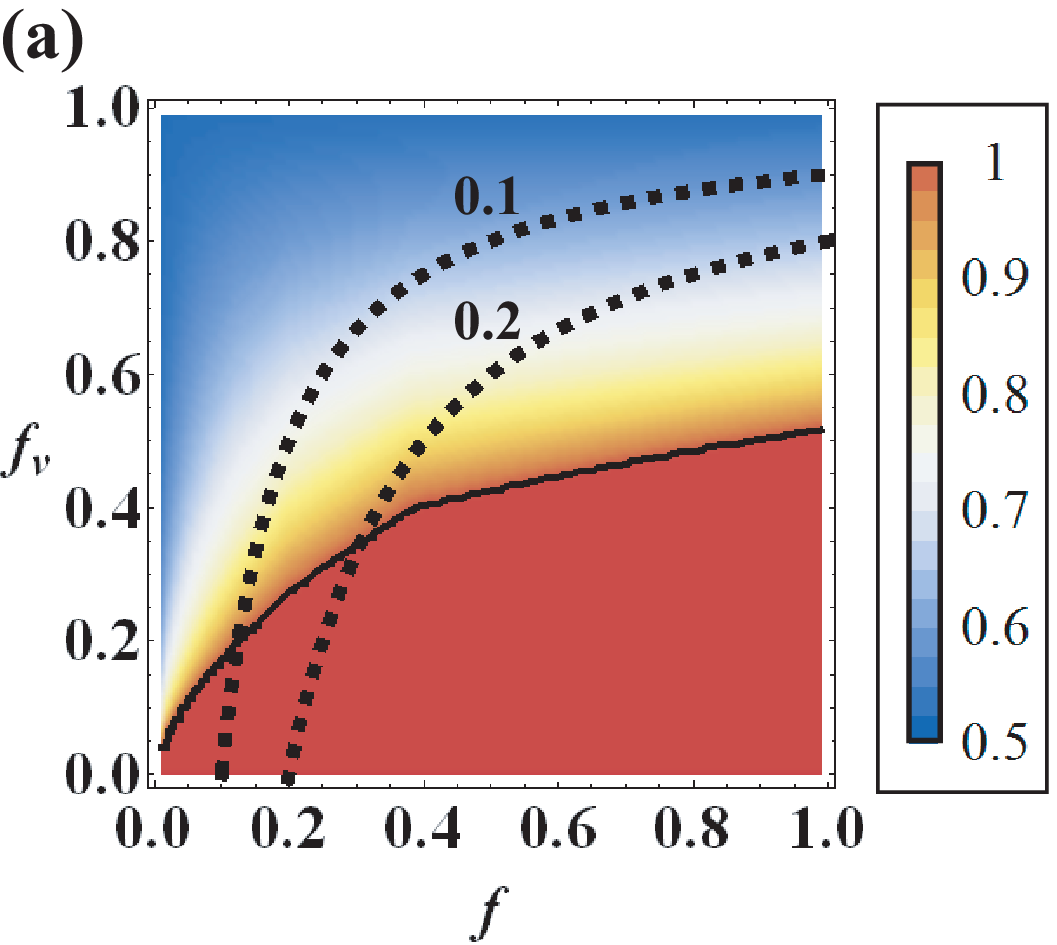}
\includegraphics[height=5cm]{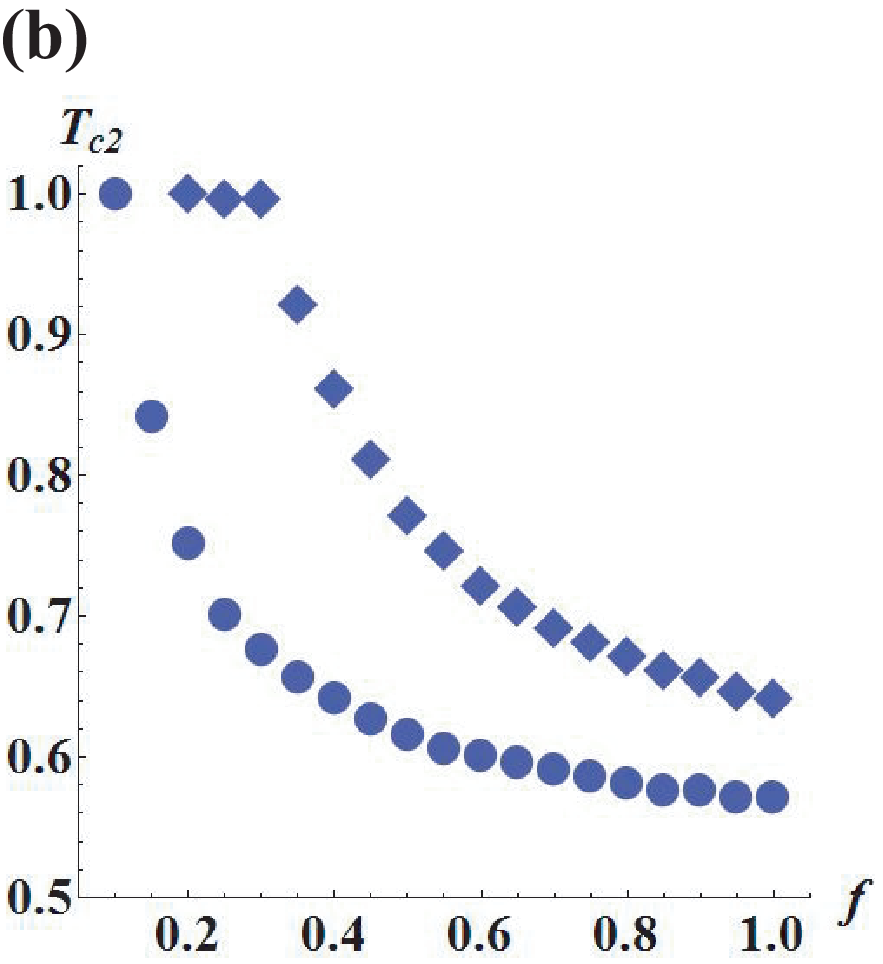}
\end{center}
\caption{Results for the SF network with the degree-based defense (case (iii)).
(a) $T_{\rm c2}$ as a function of $f$ and $f_{\rm v}$. 
The solid line indicates $T_{\rm c2}=1$.
The dashed lines indicate $f(1-f_{\rm v})=0.1$ and $f(1-f_{\rm v})=0.2$.
(b) $T_{\rm c2}$ conditioned by $f(1-f_{\rm v})=0.1$ (circles) and $0.2$ (diamonds).
}
\label{Phase-SFN-Target}
\end{figure}


In figures~\ref{Phase-ER}, \ref{Phase-SFN-Rand}, and \ref{Phase-SFN-Target}, 
$T_{\rm c1}$ and $T_{\rm c2}$ obtained from Eqs.~(\ref{1stTdef}) and (\ref{2ndTdef}) 
are shown for the three cases.

$T_{\rm c1}$ monotonically increases with $f$ and monotonically decreases with
$f_{\rm v}$ in case (i) (figure~\ref{Phase-ER}(a)).
When $T_{\rm c1}=1$ (the red region bounded by the bold line in figure~\ref{Phase-ER}(a)), 
a global infection does not occur at any infection rate.
On the basis of Eq.~(\ref{1stTdef}),
the boundary of the region $T_{\rm c1}=1$ is given by
\begin{equation}
1-f+f_{\rm v}f= \frac{\langle k \rangle}{\langle k(k-1) \rangle}.
\end{equation}
In case (ii), $T_{\rm c1}=0$ for any $f$ and $f_{\rm v}$ unless $(f,f_{\rm v})=(1,0)$.
In case (iii),
$T_{\rm c1}=0$ if $f_{\rm v}>0$ or $(f,f_{\rm v})=(0,0)$ for the following reason.
If $f_{\rm v}>0$, ${F_1^{\rm V}}^{\prime}(1)$ in Eq.~(\ref{1stTcondition})
diverges. If $f=0$, 
${F_1^{\rm U}}^{\prime}(1)$ in Eq.~(\ref{1stTcondition}) diverges (and
${F_1^{\rm V}}^{\prime}(1)=0$).

Similar to $T_{\rm c1}$,
$T_{\rm c2}$ monotonically increases with $f$ and monotonically decreases with
$f_{\rm v}$ in all the cases (figures~\ref{Phase-ER}(b), \ref{Phase-SFN-Rand}(a), 
and \ref{Phase-SFN-Target}(a)).
When $T_{\rm c2}=1$, 
the residual network is not disintegrated 
at any infection rate
(the red regions bounded by the bold line in figures~\ref{Phase-ER}(b), \ref{Phase-SFN-Rand}(a), 
and \ref{Phase-SFN-Target}(a)).

When $f_{\rm v}=0$, the SF network yields
$T_{\rm c2}=1$ for any $f>0$, either for the random defense 
(figure~\ref{Phase-SFN-Rand}(a)) and the degree-based defense
(figure~\ref{Phase-SFN-Target}(a)). 
In this situation, V nodes are never infected such that they percolate
in a SF network. This phenomenon is identical to the standard site percolation
on SF networks with $\gamma \le 3$;
randomly located nodes percolate even at an infinitesimally small occupation probability~\cite{callaway2000network,cohen2000resilience}.

The comparison of the values of $T_{\rm c2}$ among the three cases 
based on figures~\ref{Phase-ER}(b), \ref{Phase-SFN-Rand}(a), and \ref{Phase-SFN-Target}(a)
suggests the following.
When the vaccines are not efficient (i.e., $f_{\rm v} \sim 1$), 
the network is the most robust for case (i) among the three cases in the sense that
$T_{\rm c2}$ is the largest.
In this situation, $T_{\rm c2}$ for the SF network is slightly
smaller than that for the random graph (also see
figure~\ref{Comp-ER-rand}).
When $f_{\rm v}$ takes an intermediate value, the network is the
most robust for case (iii).
The network is the most fragile for case (ii)
except for fairly small $f_{\rm v}$ and $f$, in which case
the residual network mainly composed of vaccinated nodes can 
percolate on the SF network and not on the random graph.

In reality, 
there may be a trade-off between 
the required number of vaccines and the efficiency of vaccines.
Therefore, we measure
$T_{\rm c2}$ under the constraints
$f(1-f_{\rm v})=0.1$ and $f(1-f_{\rm v})=0.2$;
$f$ and $1-f_{\rm v}$ represent the number and efficiency of vaccines, respectively. 
As shown in figures~\ref{Phase-ER}(c), \ref{Phase-SFN-Rand}(b), 
and \ref{Phase-SFN-Target}(b), 
the network becomes progressively robust
as $f$ decreases (i.e., small number of vaccines)
and $f_{\rm v}$ decreases (i.e., high efficiency of vaccines)
in all the three cases.


\begin{figure}
\begin{center}
\includegraphics[width=10cm]{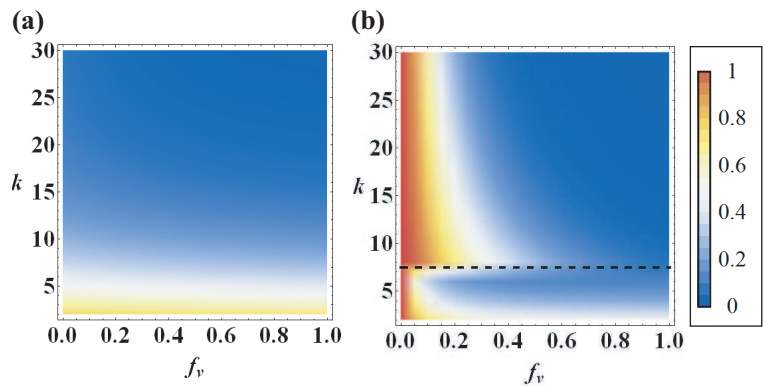}
\end{center}
\caption{Probability $U(k)$ that a node with degree $k$ in the SF network
is uninfected. (a) Random defense (case (ii)).
(b) Degree-based defense (case (iii)).
We set $T_0=0.4$ and $f=0.05$.
The dashed line in (b) indicates $k_{\rm cut}=8$.
}
\label{prevalence-ER-rand}
\end{figure}


The difference between the random and degree-based defenses
in the SF network is apparent when we look at $U(k)$, i.e., the
probability that a node with degree $k$ remains uninfected. 
In figures~\ref{prevalence-ER-rand}(a) and \ref{prevalence-ER-rand}(b),
$U(k)$ values are shown for cases (ii) (Eq.~(\ref{resultUkrandom})) 
and (iii) (Eq.~(\ref{resultUktarget})), respectively. 
As expected, 
the degree-based defense cuts down the probability of
infection for nodes with high degrees, whereas the random defense does not.
Furthermore, the indirect effect of vaccines, i.e., the suppression of
the infection probability for U
nodes, is strong under the degree-based defense. For example,
some fractions of U nodes with degree $k<k_{\rm cut}$
(below the dashed line
in figure~\ref{prevalence-ER-rand} (b)) remain uninfected.

\section{Summary}

We investigated the effect of defense strategies on the robustness of networks
against propagating attacks by extending the
generating function framework developed by
Newman~\cite{Newman-Threshold-2005PRL}. In particular, under vaccination
strategies,
we analytically obtained the second critical infection rate above which
the residual network is disintegrated.
We applied the analytical results to the three cases:
the random graph with the random defense, 
the SF network with the random defense, and the SF network
with the degree-based defense. 
The first critical point of the SF network 
depends crucially on whether the effect of the vaccines is perfect or leaky. 
Even under the degree-based defense, 
the leaky vaccines cannot prevent a global outbreak on heterogeneous networks.
We also found that, under a trade-off between the number and efficiency of vaccines,
it is better to administer a small number of vaccines with high efficiency than vice versa
to prevent the residual network from being disintegrated.

\ack

NM
acknowledges financial support by the Grants-in-Aid for Scientific Research
(No.\ 20760258) from MEXT, Japan. 
This research is also partially supported by Aihara Innovative Mathematical
Modelling Project, the Japan Society for the Promotion of Science
(JSPS) through the "Funding Program for World-Leading Innovative R\&D
on Science and Technology (FIRST Program)," initiated by the Council
for Science and Technology Policy (CSTP).


\end{document}